\documentclass{elsart1p}
\usepackage{graphicx}
\usepackage{amssymb}
\usepackage{epsfig}

\begin{document}
\begin{frontmatter}
%
%
%
%
%
\title{Elliptic flow of light nuclei in Au+Au collisions at
  $\sqrt{s_{NN}}$ = 200 GeV }
%
%

\author{Chitrasen Jena (for the STAR Collaboration)}

\address{Institute of Physics, Bhubaneswar-751005, INDIA.}

\begin{abstract}
  We present the elliptic flow ($v_{2}$) of light nuclei at
  mid-rapidity in Au+Au collisions at $\sqrt{s_{NN}}$ = 200 GeV.  The
  results are measured in the STAR experiment at RHIC. The $v_{2}$
  measurement for light nuclei as a function of transverse momentum
  $p_{T}$ is found to follow an approximate atomic mass number ($A$)
  scaling. We compare the measured nuclei $v_{2}$ to results from a
  dynamical coalescence model calculation. The $v_{2}$ values for
  light nuclei are further scaled to the number of constituent quarks
  (NCQ) of their constituent nucleons and are consistent with NCQ
  scaled $v_{2}$ for baryons and mesons. The dominance of partonic
  collectivity in the transverse expansion dynamics in these
  collisions naturally produces such a consistent picture. Similar to other
  hadrons, an increase of $p_{T}$-integrated $v_{2}$ scaled by the
  participant eccentricity as a function of collision
  centrality has been observed.

\end{abstract}

%

\end{frontmatter}

\bibliographystyle{unsrt}
\def\Journal#1#2#3#4{{#1} {\bf #2}, #3 (#4)}

\def\NCA{ Nuovo Cimento}
\def\NIM{ Nucl. Instrum. Methods}
\def\NIMA{{ Nucl. Instrum. Methods} A}
\def\NPB{{ Nucl. Phys.} B}
\def\NPA{{ Nucl. Phys.} A}
\def\PLB{{ Phys. Lett.}  B}
\def\PRL{ Phys. Rev. Lett.}
\def\PR{ Phys. Rev.}
\def\PRC{{ Phys. Rev.} C}
\def\PRD{{ Phys. Rev.} D}
\def\ZPC{{ Z. Phys.} C}
\def\JPG{{ J. Phys.} G}
\def\EJC{{ E. J. Phys.} C}
\def\CTP{ Commun. Theor. Phys.}

\def\st{\scriptstyle}
\def\sst{\scriptscriptstyle}
\def\mco{\multicolumn}
\def\epp{\epsilon^{\prime}}
\def\vep{\varepsilon}
\def\ra{\rightarrow}
\def\ppg{\pi^+\pi^-\gamma}
\def\vp{{\bf p}}
\def\ko{K^0}
\def\kb{\bar{K^0}}
\def\al{\alpha}
\def\ab{\bar{\alpha}}
\def\bea{\begin{eqnarray}}
\def\eea{\end{eqnarray}}
\def\CPbar{\hbox{{\rm CP}\hskip-1.80em{/}}}

\section{Introduction}
\label{}
One of the most important experimental findings at RHIC has been the
signature of coalescence as the mechanism of particle production. The
differences in baryons and mesons at the intermediate transverse
momentum ($1.5 < p_{T} < 5$ GeV/c) for observables like the nuclear modification
factor and the elliptic flow parameter have been attributed to be the
signatures of quark coalescence as a mechanism of hadron
production~\cite{fries}. However, it is experimentally difficult to
study how local correlations and energy/entropy play a role in
coalescence at the partonic level since the constituents are not
directly observable.  
In relativistic heavy-ion collisions, light nuclei and anti-nuclei are
formed through coalescence of nucleons and
anti-nucleons~\cite{final_state, coalescence1}. The binding energy for the light
nuclei are small, hence this formation process can only happen at a
late stage of the evolution of the system when interactions between
nucleons and other particles are weak. This process is called
final-state coalescence~\cite{final_state, BA_ref}. The coalescence
probability is related to the local nucleon density. Therefore, the
production of light nuclei provides an interesting tool to measure collective
motion and freeze-out properties. The advantage of
nucleons over the partonic coalescence phenomena is that both the
nuclei and the constituent nucleon space-momentum distributions are
measurable quantities in heavy-ion collisions. By studying the
elliptic flow of nuclei and comparing to those of their constituents
(nucleons), we will have a better understanding of coalescence process
for hadronization.

\section{Experiment and analysis}
\label{}

\begin{figure}
\centering
\begin{tabular}{cc}
\epsfig{file=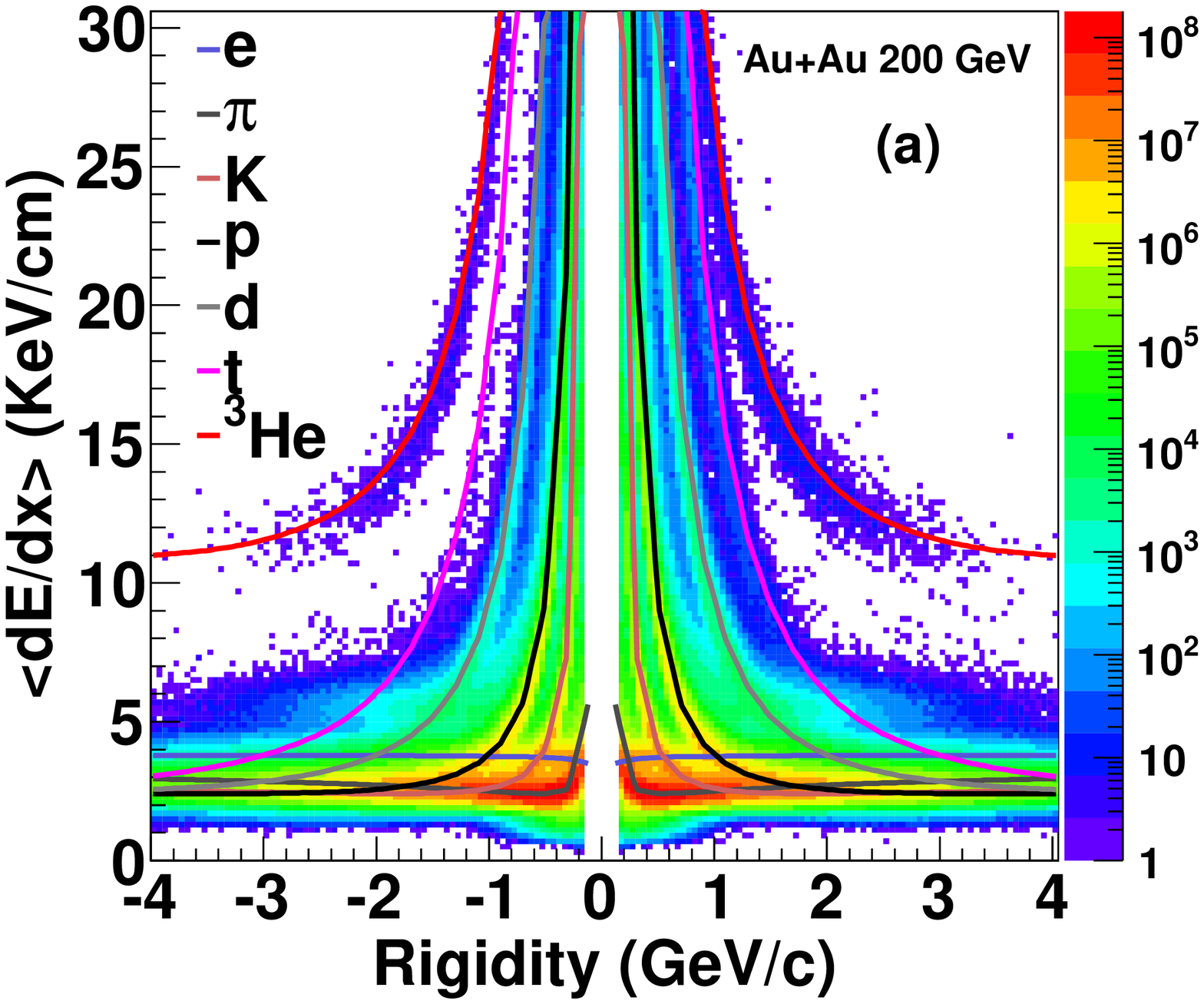,angle=0,width=0.42\linewidth,clip=} &

\epsfig{file=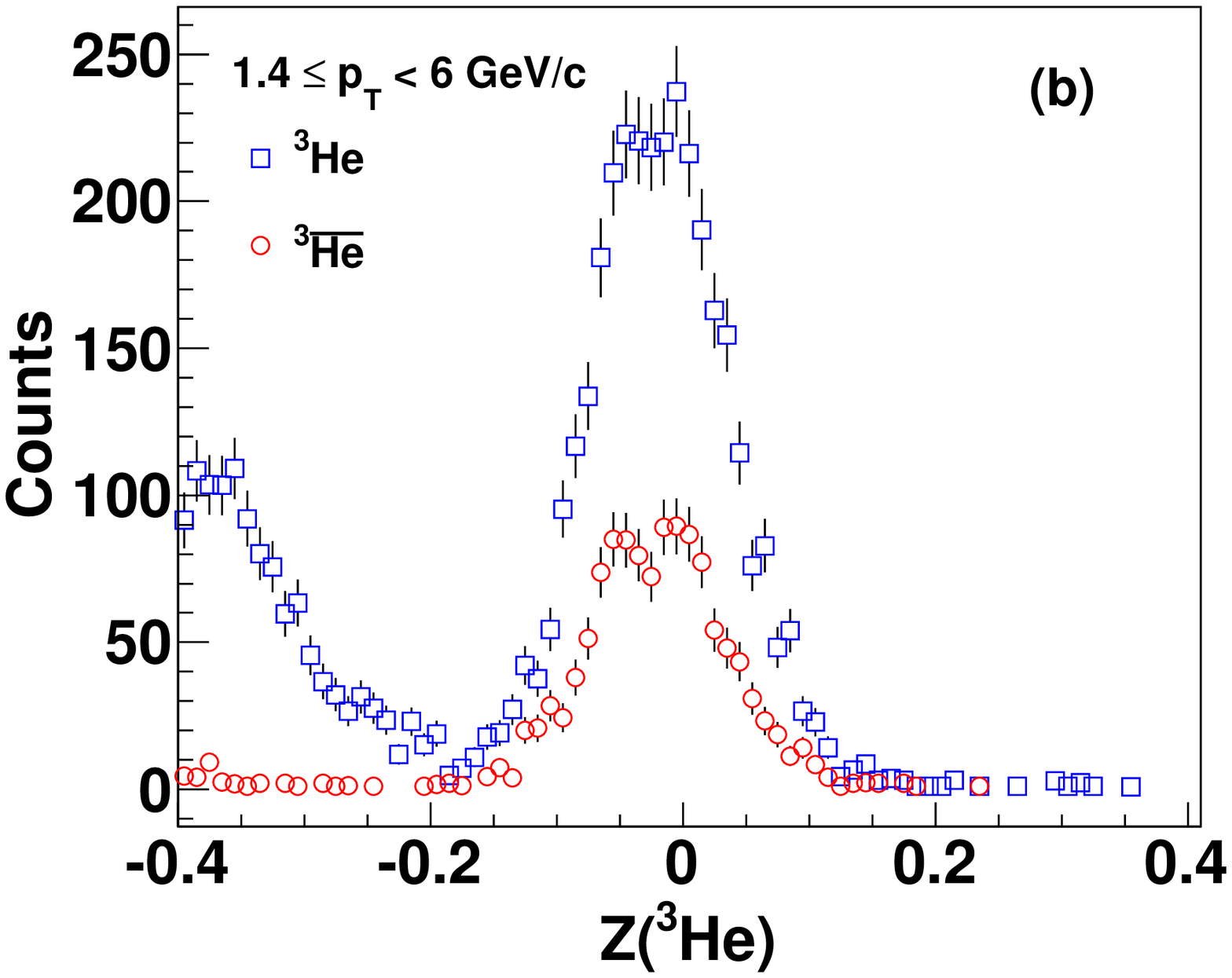,angle=0,width=0.45\linewidth,clip=} \\
\end{tabular}
\caption{(a) TPC $dE/dx$ as a function of rigidity. Lines are
  exepected $dE/dx$  values for different charged tracks
  predicted by the Bichsel function  \cite{bichsel}. 
(b)  $Z$ distribution of $^{3}He$ (open square) and $\overline{^{3}He}$ (open
circle). See the text for details.} 
\label{Tech_figure}
\end{figure}

The data presented here are obtained from Au+Au collisions at
$\sqrt{s_{NN}}$ = 200 GeV with the STAR detector at RHIC in the year
2007. STAR's main Time Projection Chamber (TPC)~\cite{STAR} was used
for tracking and identification of charged particles. A data sample of
62 million minimum-bias triggered events (0-80\% centrality) is used for
this analysis. Figure 1 presents the particle identification
techniques and methods. Panel (a) shows the ionization energy loss
($dE/dx$) of charged tracks as a function of rigidity
($rigidity=momentum/charge$) measured by the TPC at $-1<\eta<1$. Panel
(b) shows $Z$ ($Z=\log((dE/dx)|_{measure}/(dE/dx)|_{predict})$)
distribution for $^{3}He$ and $\overline{^{3}He}$ signals for $1.4 \le
p_{T} < 6$ GeV/c, where
$(dE/dx)|_{predict}$ is the $dE/dx$  value predicted by the
Bichsel function~\cite{Ming, bichsel}.  After tight track quality
selections,  the $^{3}He(\overline{^{3}He})$ signals are essentially
background free. We derive the yields by counting
$^{3}He(\overline{^{3}He})$ candidates with $|Z(^{3}He)| < 0.2$.

\section{Results}
\label{}

\begin{figure}
\centering
\begin{tabular}{cc}
\epsfig{file=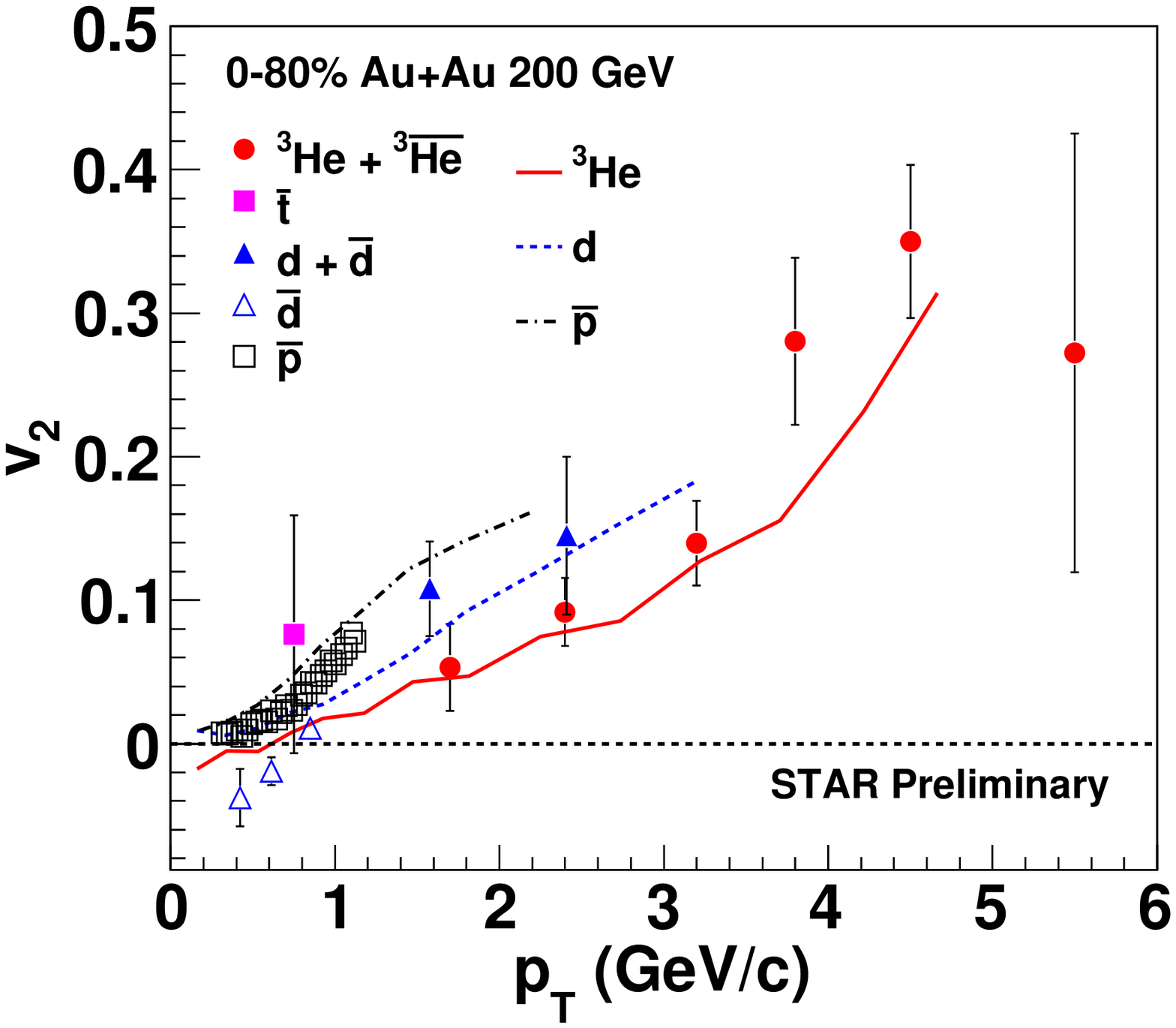,angle=0,width=0.4\linewidth,clip=} &

\epsfig{file=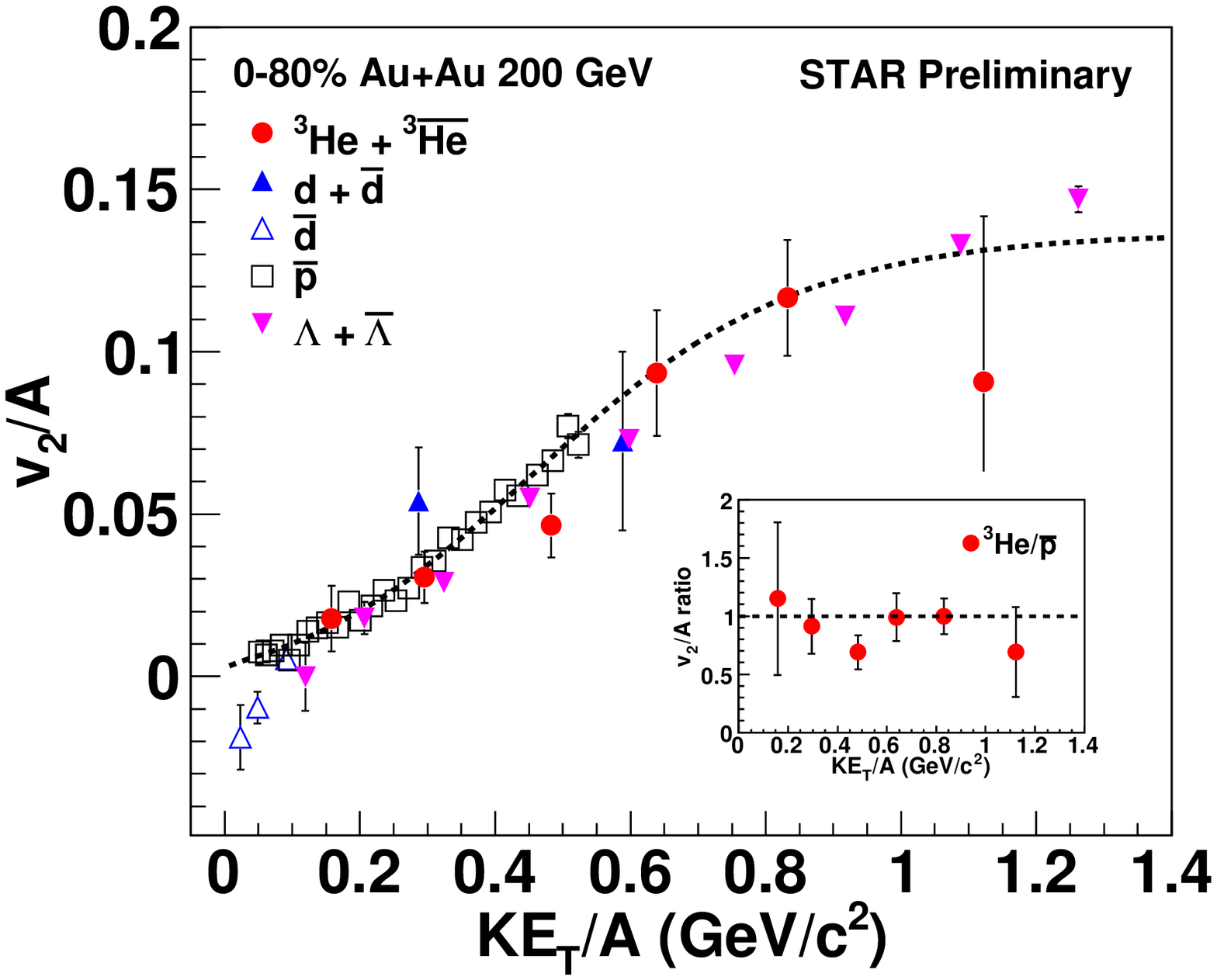,angle=0,width=0.41\linewidth,clip=} \\
\end{tabular}
\caption{ Left panel: $v_{2}$ as a function of $p_{T}$  for $\overline{t}$
  and $^{3}He $+$ \overline{^{3}He}$ from 0-80\% of the collision
  centrality. $d(\overline{d})$ and $\overline{p}$ $v_{2}$ are shown in the
  plot as a comparison \cite{run4_paper} . The $v_{2}$ calculations from
  dynamical coalescence model are shown in different lines. 
Right panel: $d(\overline{d})$ and $^{3}He $+$ \overline{^{3}He}$  $v_{2}$ as a
function of $KE_{T}$, both $v_{2}$ and $KE_{T}$ have been scaled by
A.  $\overline{p}$ (open square) and $\Lambda+\overline{\Lambda}$
(solid inverted triangle) $v_{2}$ are shown in the plot as a
comparison. The dotted line is a fit to the $v_{2}$ of
$\overline{p}$. The insert is the ratio of $^{3}He$+$\overline{^{3}He}$ to $\overline{p}$}.
\label{flow_fig_1}
\end{figure}

\begin{figure}
\centering
\begin{tabular}{cc}
\epsfig{file= 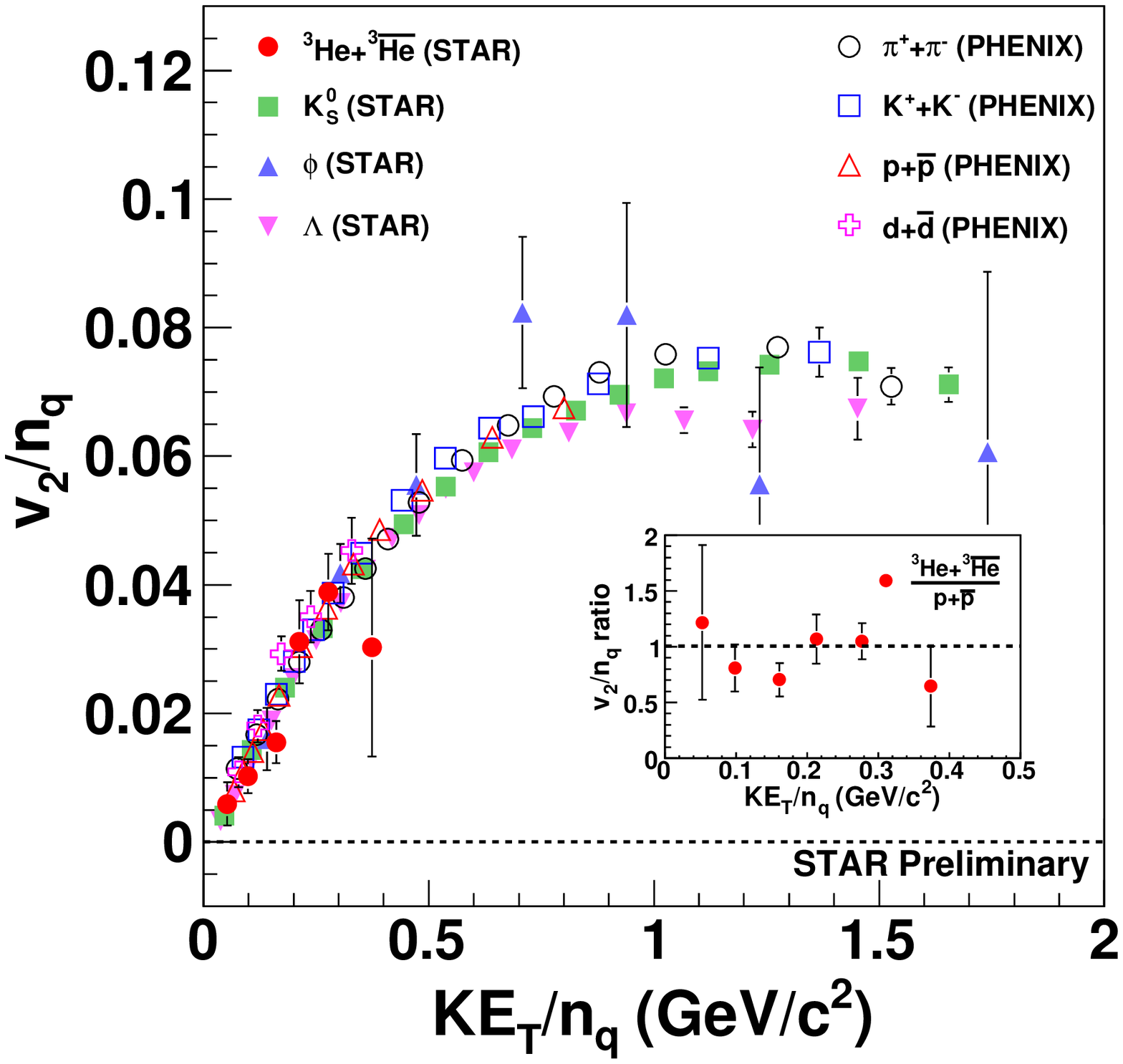,angle=0,width=0.48\linewidth,clip=} &

\epsfig{file=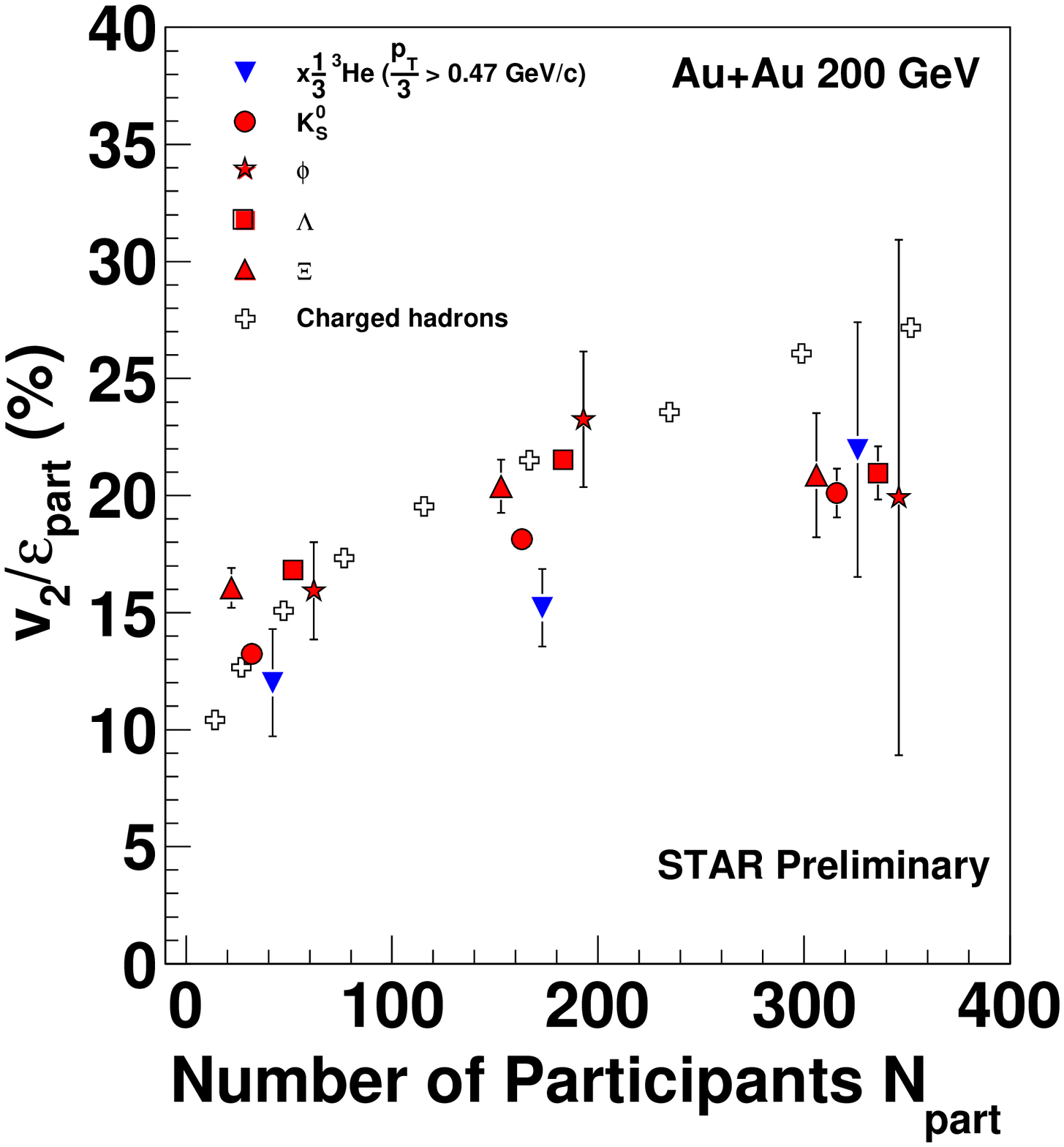,angle=0,width=0.44\linewidth,clip=} \\
\end{tabular}
\caption{ Left panel:  $v_{2}/n_{q}$ as a function of $KE_{T}/n_{q}$
 for different particles. The insert is the ratio of
 $^{3}He$+$\overline{^{3}He}$ to $p+ \overline{p}$. Right panel:
 $v_{2}/\varepsilon$ as a function of centrality for different particles,
 the $^{3}He $ value is scaled down by a factor of 3.}
\label{flow_fig_2}
\end{figure}

The elliptic flow parameter, $v_{2}$, is the second order Fourier
coefficient of the azimuthal distribution of the produced nuclei
relative to the reaction plane of the initial nucleus-nucleus
collision. The event-plane method was used to obtain the $v_{2}$ of
nuclei~\cite{v2method}, with the event plane resolution used for correction (calculated using the
sub-event method~\cite{v2method}) being 73\% for minimum bias triggered
events. In the following discussions only statistical errors will be
shown.
 
The left panel of Fig.~\ref{flow_fig_1} shows $v_{2}$ as a function of
$p_{T}$ for $\overline{t}$ and $^{3}He+\overline{^{3}He}$ in
minimum-bias (0-80\% centrality) collisions. The $v_{2}$ of $\overline{t}$ is
shown only for $0.3 < p_{T} < 1.2$ GeV/c. In the future, high statistics
data sets and good particle identification using STAR TPC and TOF
(Time-of-Flight) will allow us to study the $p_{T}$ dependence of $v_{2}$  for
$t(\overline{t})$. The $v_{2}$ of $\overline{p}$ and
$d(\overline{d})$ measured by the STAR \cite{run4_paper} are also
shown for the comparison.  The $v_{2}$ of $p$($\overline{p}$),
$d$($\overline{d}$) and $^{3}He$($\overline{^{3}He}$)  are well
described by the dynamical coalescence model \cite{dynaCoal}. In this model, the
probability for producing a cluster of nucleons ($d$, $t$ and $^{3}He$) is
determined by the overlap of its Wigner phase-space density with the
nucleon phase-space distributions at freeze-out. To determine the
Wigner phase-space densities of the $d$, $t$ and $^{3}He$, we take
their hadron wave functions to be those of a spherical harmonic
oscillator. The coordinate and momentum space distributions of hadrons
at freeze-out are obtained from AMPT model calculation~\cite{AMPT}.

The results with both $v_{2}$ and transverse kinetic energy $KE_{T} =
m_{T} - m$ , where $m_{T} = \sqrt{p_{T}^{2} + m^{2}}$ scaled by $A$
are shown in the right panel of Fig.~\ref{flow_fig_1}.  As a comparison, the $\overline{p}$ and
$\Lambda+\overline{\Lambda}$ $v_{2}$ \cite{run4_paper} are superimposed
on the plot. The data suggest that the $^{3}He$($\overline{^{3}He}$),
$d$($\overline{d}$) and baryon $v_{2}$ seem to follow the $A$ scaling
within errors, indicating that the light nuclei are formed through
the coalescence of nucleons just before thermal freeze-out. 

The left panel of Fig.~\ref{flow_fig_2} shows $v_{2}/n_{q}$ as
a function of $KE_{T}/n_{q}$ for different hadrons inluding the nuclei
and anti-nuclei, where $n_{q}$ is the number of constituent
quarks. The value of $n_{q}$ used for $d$($\overline{d}$) and
$^{3}He$($\overline{^{3}He}$) are 6 and 9 respectively, to account for
their composite nature. The scaled results for $v_{2}$ versus
$KE_{T}$ for the light nuclei and anti-nuclei are consistent with the
experimentally observed NCQ scaling of $v_{2}$ for baryons and mesons
\cite{ncq_PHENIX}. This is also consistent with the picture that
partonic collectivity dominates the transverse expansion dynamics of the
nucleus-nucleus collisions at RHIC. The right panel of
Fig.~\ref{flow_fig_2}  shows the centrality dependence of the ratio
of the integrated $v_{2}$ over the eccentricity
($v_{2}/\varepsilon_{part}$) for charged hadrons, $K_{s}^{0}$, $\phi$
meson, $\Lambda$, $\Xi$ and $^{3}He$.  The $v_{2}$ of $^{3}He$ is
integrated over the measured $p_{T}$ range (i.e. $p_{T} > 1.4$ GeV/c )
and scaled down by a factor of 3.  This ratio ($v_{2}/\varepsilon_{part}$) to some extent reflects
the strength of the collective expansion. For more central collision,
the larger value of this ratio indicates a stronger collective expansion.

\section{Summary}
\label{}
We have measured the $v_{2}$ of light nuclei using the particle
identification capability of the STAR TPC detector. The $v_{2}$ values
of the light nuclei when scaled by atomic mass number $A$, follows
the baryon $v_{2}$, which supports the idea of the light nuclei
formation through the coalescence of nucleons just before thermal
freeze-out. The $v_{2}$ of light nuclei from the dynamical coalescence 
model agree well with the data. The $v_{2}$ as a function of the 
transverse kinetic energy  follows a scaling with  the number of
constituent quarks $n_{q}$.



\begin{thebibliography}{00}

\bibitem{fries} R. Fries {\it et al.}, Ann.Rev.Nucl.Part.Sci. {\bf58},177 (2008).

\bibitem{final_state}H.H. Gutbrod {\it et al.}, \Journal{\PRL}{37}{667}{1976}.

\bibitem{coalescence1}R. Scheibl and U. Heinz, \Journal{\PRC}{59}{1585}{1999}. 

\bibitem{BA_ref}S.T. Butler and C.A. Pearson, \Journal{\PR}{129}{836}{1963}.

\bibitem{quarkcoalescence}
D. Molnar and S. A. Voloshin, \Journal{\PRL}{91}{092301}{2003}; 
R. C. Hwa and C. B. Yang, \Journal{\PRC}{67}{064902}{2003};
R. J. Fries {\it et al.} \Journal{\PRL}{90}{202303}{2003}. 

\bibitem{coal_dyn}Y.S. Oh and C.-M. Ko, \Journal{\PRC}{76}{054910}{2007}. 

\bibitem{generalv2}
J.-Y. Ollitrault, \Journal{\PRD}{46}{229}{1992};
H. Sorge, \Journal{\PRL}{82}{2048}{1999}; 
K. H. Ackermann {\it et al.} (STAR Collaboration), \Journal{\PRL}{86}{402}{2001}.

\bibitem{STAR}K.H. Ackermann {\it et al.} (STAR Collaboration), \Journal{\NIMA}{499}{624}{2003}.
\bibitem{Ming}M. Shao {\it et al.}, \Journal{\NIMA}{558}{419}{2006}.
\bibitem{bichsel}H. Bichsel, \Journal{\NIMA}{562}{154}{2006}. 

\bibitem{v2method}A. M. Poskanzer and S. A. Voloshin, \Journal{\PRC}{58}{1671}{1998}.
\bibitem{run4_paper}B.I. Abelev {\it et al.} (STAR Collaboration), arXiv$:$nucl-ex/0909.0566.
\bibitem{dynaCoal} S. Zhang  {\it et al.} \Journal{\PLB}{684}{224}{2010}.
\bibitem{AMPT}Z.W. Lin, C.M. Ko, B.A. Li, B. Zhang, and S. Pal,  \Journal{\PRC}{72}{064901}{2005}.
\bibitem{ncq_PHENIX}A. Adare {\it et al.} (PHENIX Collaboration), \Journal{\PRL}{98}{162301}{2007}.


\end{thebibliography}
\end{document}